\documentclass[aps,prd,groupedaddress,showpacs]{revtex4}
\usepackage[parfill]{parskip}    
\usepackage{graphicx}
\usepackage{amssymb}
\usepackage{epstopdf}
\usepackage{tcolorbox}
\usepackage{color}
\usepackage{float}
\usepackage{ulem}
\usepackage{graphicx}
\usepackage{pdfpages}
\usepackage{amsmath}
\usepackage[colorlinks=true,citecolor=blue,urlcolor=magenta,breaklinks]{hyperref}

\begin{document}

\title{The four-components link invariant in the framework of topological field theories}

\author{M. Anda.}
\email{mandac@estud.usfq.edu.ec}
\affiliation{Departamento de F\'isica, Colegio de Ciencias e Ingenier\'ia, Universidad San Francisco de Quito,  Quito 170901, Ecuador\\}

\author{E. Fuenmayor }
\email{ernesto.fuenmayor@ciencs.ucv.ve}
\affiliation{Centro de F\'isica Te\'orica y Computacional,\\ Escuela de F\'isica, Facultad de Ciencias, Universidad Central de Venezuela, Caracas 1050, Venezuela\\}

\author{L. Leal }
\email{lleal@fis.ciens.ucv.ve}
\affiliation{Centro de F\'isica Te\'orica y Computacional,\\ Escuela de F\'isica, Facultad de Ciencias, Universidad Central de Venezuela, Caracas 1050, Venezuela\\}

\author{E. Contreras }
\email{econtreras@usfq.edu.ec (corresponding author)}
\affiliation{Departamento de F\'isica, Colegio de Ciencias e Ingenier\'ia, Universidad San Francisco de Quito (USFQ),  Quito 170901, Ecuador\\}

\begin{abstract}
In this work, we undertake a perturbative analysis of the topological non-Abelian Chern-Simons-Wong model with the aim to explicitly construct the second-order on-shell action. The resulting action is a topological quantity depending solely on closed curves, so it correspond to an analytical expression of a link invariant. Additionally, we construct an Abelian model that reproduces the same second-order on-shell action as its non-Abelian Chern-Simons-Wong counterpart so it functions as an intermediate model, featuring Abelian fields generated by currents supported on closed paths. By geometrically analyzing each term, we demonstrate that this topological invariant effectively detects the knotting of a four-component link.
\end{abstract}

\maketitle

\section{Introduction}\label{intro}

The exploration of quantum Chern-Simons (CS) theory  and its connection to Knot Theory (KT) is a vibrant area of investigation in both Mathematics and Theoretical Physics.
The connection between field and knot theories was established by E. Witten, who found that the vacuum expectation value of the Wilson Loop (WL) produces a knot invariant closely connected to the Jones polynomials \cite{witten89,jones,guadagnini}. Then it was found that the knot invariants derived from a perturbative expansion of the WL average \cite{guadagnini} offer solutions to the constraints of Ashtekar Quantum Gravity \cite{ashtekar86} in the Loop Representation \cite{dibartolo99,rovelli90}. Motivated by these results, a large number of authors have contributed significantly to the better understanding and use of knot invariants in modern physics \cite{rolsfen,mona,L1,L2,L3,L4,fadd,rozanski,L5,L6,china1,china2,resh1,resh2,EF}. \\

In Ref. \cite{leal2002}, the author demonstrates that the interplay between CS and KT is manifested also at the classical level as
the action  of the theory \cite{wong,bala} retains its diffeomorphism-invariant character when it is evaluated on-shell (this result can be rigorously proven and generalized to situations where the symmetry group is other than the group of diffeomorphisms of the base manifold \cite{leal3}). Hence, the on-shell action of the CS theory coupled in a suitable manner to particles should yield link invariant (LIs) in three dimensional spacetimes, just as the vacuum expectation value of the Wilson-Loop does within the quantum field approach.  Then it is clear that the on-shell action of this model is expected to result in analytical representations for LIs associated with the trajectories of the particles. However, given the non-linearity of the system, which hinders obtaining exact solutions for the equations of motion, the author devised a perturbative approach for their solution. In particular, in \cite{leal2002} it was obtained the first two contributions to the on-shell action and were identified as the Second and Third Milnor Linking Coefficients \cite{milnor}. This family of LIs, discovered by Milnor \cite{milnor} and characterized by increasing complexity, includes the Gauss Linking Number, which coincides with the Second Coefficient and the LI of the Borromean rings which coincides with the third \cite{leal2002}.\\ 

A natural question arises concerning the possibility of an intermediate scenario lying between the Abelian and non-Abelian cases. To be more precise: is there any topological field theory, distinct from Abelian CS, capable of providing exact analytical expressions for LIs other than the Gauss Linking Number (GLN)? This question holds not only theoretical significance but also gains practical importance due to the growing interest in describing phenomena featuring closed lines as essential structures, such as vortices and defects in condensed matter or fundamental physics, loops in gauge theories and quantum gravity, polymer entanglements, among other examples. Consequently, the availability of new topological theories, extending beyond Abelian CS theory and its associated GLN, yet avoiding the challenges posed by non-Abelian theories, could prove beneficial. The answer to this question is affirmative, and it was successfully addressed in \cite{leal-intermedia}, where the authors proposed an Abelian action that precisely recovers the second term of the expansion developed in 
\cite{leal2002}. \\

The goal of this work is twofold. On the one hand, we will explicitly calculate the second order perturbation of the on-shell action as a continuation of \cite{leal2002} with the aim of subsequently proposing an Abelian intermediate action that reproduces the same result following \cite{leal-intermedia}. On the other hand, we will demonstrate that our result is associated with the linking properties of a four-component link by using a practical language, which facilitates the calculations and interpretation \cite{dibartolo}. We also discuss the consistency
conditions that gauge invariance imposes over the classical equations of motion, and their
relationship with the definition of the link-invariants.\\

This work is organized as follows. In the next section, we introduce the basic concepts of the Chern-Simons-Wong (CSW) model. In particular, we reproduce the on-shell action to $p$-th order reported in \cite{leal2002} and explicitly write down the action to second order. In Section \ref{ia}, we propose an intermediate action and demonstrate that it exactly reproduces the second-order action from the perturbative development of the non-Abelian theory. Then, in Section \ref{gi}, we express each term of the invariant in a convenient way and demonstrate its association with the knotting of a four-component link. Finally, the last section is dedicated to the conclusions of the work.

\section{Chern-Simons-Wong: a perturbative approach}\label{cst}

The action of the CSW theory is given by
\begin{equation}
    S=S_{C S}+S_{i n t},
    \label{tcs}
\end{equation}
where $S_{CS}$ is the $SU(N)$ CS term
\begin{equation}
    S_{C S}=-\Lambda^{-1} \int d^3 x \epsilon^{\mu \nu \rho} \operatorname{Tr}\left(A_\mu \partial_\nu A_\rho+\frac{2}{3} A_\mu A_\nu A_\rho\right),
\end{equation}
and $S_{int}$ corresponds to the field-particle interaction of $n$ non-dynamical Wong particles (i.e.,
classical particles with cromo-electrical charge) \cite{wong} given by
\begin{equation}
    S_{i n t}=\sum_{i=1}^n \int_{\gamma_i} d \tau \operatorname{Tr}\left(K_i g_i^{-1}(\tau) D_\tau g_i(\tau)\right),
\end{equation}
with  $\Lambda$ is a constant, $g_i(\tau)$ are matrices (belonging to the $SU(N)$ in the fundamental representation) associated with the internal degrees of freedom of the particles, $D_\tau g_i(\tau)$ is the covariant derivative of $g_i(\tau)$ along the worldline of the $i$-th particle $\gamma_i$
\begin{equation}
    D_\tau g_i(\tau)=\dot{g}_i(\tau)+A_i(\tau) g_i(\tau),
\end{equation}
and $K_i = K_i^a T^a$ is a constant element in the algebra related to the initial value of the chromoelectric charge $I_i(\tau)$ defined by
\begin{equation}\label{definitionI}
    I_i(\tau) \equiv g_i(\tau) K_i g_i^{-1}(\tau).
\end{equation}

The following conventions and definitions for the $N^{2}-1$ generators $T^a$ of $su(N)$ and the gauge field $A_\mu$ are used throughout the paper:
\begin{eqnarray*}
   Tr(T^a T^b)&=&- \frac12\delta^{ab}\\
  \left[ T^{a} , T^{b} \right] &=& f^{abc} T^c\\
    A_{\mu}&=&A_{\mu}^a T^a\\
    A_i&=&A_{\mu}(z_i (\tau)) \dot{z}_i^{\mu}.
\end{eqnarray*}
As dynamical variables we take the potentials $A_{\mu}$ and the matrices $g_i (\tau)$ associated with the internal degrees of freedom of the Wong particles \cite{wong}. Besides, we consider that the trajectories $z_i ^{\mu} (\tau)$ are externally given, and assume that the curves $\gamma_i$ drawn by these particles are closed curves in Euclidean three space whose linking properties we are going
to explore. \\

At this point a couple of comments are in order. First, note that the model is invariant under the following transformations
\begin{eqnarray*}
   A_\mu^{\Omega}&=&\Omega^{-1} A_\mu \Omega+\Omega^{-1} \partial_\mu \Omega\\
    K_i^{\Omega}&=&K_i\\
    g_i^{\Omega}&=&\Omega^{-1} g_i\\
    I_i^{\Omega}&=&\Omega^{-1} I_i \Omega.
\end{eqnarray*}
Second, it is worth mentioning that we could consider dynamic trajectories, but for our purposes, this is not convenient, as we want to treat the particle world lines as external objects, focusing solely on their link properties as in \cite{leal2002}.\\

In what follows, we will focus in obtaining the equations of motions from the action (\ref{tcs}). Varying the action with respect to $A_{\mu}$ we obtain the field equations of the CS theory with matter
\begin{equation}
    \varepsilon^{\mu \nu \rho} F_{\nu \rho}=\Lambda J^\mu
    \label{ec1}
\end{equation}
where
\begin{equation}\label{corriente}
    J^\mu(x)=\sum_{i=1}^n \int_{\gamma_i} d \tau \dot{z}_i^\mu(\tau) I_i(\tau) \delta^3\left(x-z_i(\tau))\right.
\end{equation}
and 
\begin{equation}
    F_{\mu \nu}=\partial_\mu A_\nu-\partial_\nu A_\mu+\left[A_\mu, A_\nu\right].
\end{equation}
Next, to varying the action with respect to the $SU(N)$ elements $g(\tau)$, it is necessary to separate the independent degrees of freedom which can be achieved by expressing the group elements as
\begin{equation}
    g_i=g_i\left(\xi_i\right)=e^{\xi_i^a T^a}
\end{equation}
and then taking variations respect to the $N^{2}-1$ independent parameters $\xi^{a}$ \cite{bala}. Such variations lead to the Euler-Lagrange equations
\begin{equation}\label{e-l}
    \frac{\partial L}{\partial \xi_i^a}-\frac{d}{d \tau}\left(\frac{\partial L}{\partial \dot{\xi}_i^a}\right)=0,
\end{equation}
where
\begin{equation}
L  \equiv \sum_{i=1}^n \operatorname{tr}\left(K_i g_i^{-1}(\tau) D_\tau g_i(\tau)\right),
\end{equation}
form where, it can be seen that equations (\ref{e-l}) are equivalent to the gauge-covariant conservation of the non-Abelian charge of each particle along its world line \cite{bala,leal2002}
\begin{equation}\label{con}
    D_\tau I_i \equiv \dot{I}_i+\left[A_i, I_i\right]=0.
\end{equation}
It is possible to formally integrate this equation to obtain the general solution
\begin{equation}\label{conservation}
    I_i(\tau)=U_i(\tau) I_i(0) U_i^{-1}(\tau),
\end{equation}
where
\begin{equation}\label{exponencial}
    U_i(\tau)=\mathbf{T} \exp \left\{-\int_0^\tau A_i\left(\tau^{\prime}\right) d \tau^{\prime}\right\},
\end{equation}
is the time-ordered exponential of $A_i(\tau)$ along $\gamma_i$. Plugging (\ref{exponencial}) into (\ref{ec1}) yields an equation for the gauge potentials in terms of the curves $\gamma_i$ (provided that suitable boundary conditions are given). Then, inserting the solution to this equation into the action (\ref{tcs}) one would finally obtain a functional $S(\left[\gamma \right];\Lambda)$ that only depends on the curves $\gamma$ and the coupling constant $\Lambda$.\\

It can be easily shown that the equation of motions are consistent whenever (\ref{conservation}) is satisfied. Indeed,
by taking the covariant derivative of equation (\ref{ec1}) we obtain
\begin{equation}
    \epsilon^{\mu \nu \rho} D_\mu F_{\nu \rho}=\Lambda D_\mu J^\mu,
\end{equation}
which leads to
\begin{equation}
    D_\mu J^\mu=0,
\end{equation}
after using the Bianchi identities for the Yang-Mills field \cite{jack} $\epsilon^{\mu \nu \rho} D_\mu F_{\nu \rho}=0$. Now, by using (\ref{corriente}) the above condition can be written as
\begin{equation}
\sum_i \int d \tau  \left(\frac{\partial}{\partial x^\mu} \delta^3\left(x-z_i(\tau)\right) \dot{z}_i^\mu(\tau) I_i(\tau)
+  \delta^3\left(x-z_i(\tau)\right) \dot{z}_i^\mu(\tau)\left[A_\mu(x), I_i(\tau)\right]\right)=0,
\label{ec2}
\end{equation}
form where
\begin{equation}
    \sum_i \int_0^T d \tau \delta^3\left(x-z_i(\tau)\right) D_\tau I_i=0,
\end{equation}
which is satisfied as $D_\tau I_i = 0$.\\

 We emphasize again that the action is topological as it does not depend on the metric for its construction and remains invariant under general coordinate transformations which implies that the on-shell action, $S_{OS}$, is a topological quantity too. In this sense, if we were able to solve the equations of motion (\ref{ec1}) under appropriate boundary conditions, we would find that the on-shell action depends solely on the curves $\gamma_{i}$. Given its topological nature, it should be an expression that only encodes information about how the curves are linked to each other, in other words, a LI. However, since the equation cannot be solved exactly, we can address the problem perturbatively as we will demonstrate in what follows by reviewing the approach outlined in \cite{leal2002}.\\
 
As it is shown in \cite{leal2002},
the interaction terms vanishes on-shell. Indeed, from (\ref{definitionI}) and (\ref{conservation}) we have
\begin{equation}
    \begin{aligned}
I(\tau) & =U(\tau) g(0) K g^{-1}(0) U^{-1}(\tau) \\
& =g(\tau) K g^{-1}(\tau).
\end{aligned}
\end{equation}
Now, as $g(\tau) = U(\tau) g(0)$ we obtain
\begin{equation}
    D_\tau g(\tau)=0,
\end{equation}
so that
\begin{equation}
S_{\text {int.OS}}=0 \text {. }
\end{equation}
Therefore, it only remains to consider $S_{CS}$ on-shell. To proceed further, it is convenient to express Eq. (\ref{con}) as
\begin{equation}
\frac{d I_i^a(\tau)}{d \tau}+\Lambda R_i^{a c}(\tau) I_i^c(\tau)=0,
    \label{ec5}
\end{equation}
where
\begin{eqnarray*}
    R_i^{a c} &\equiv &f^{a b c} \dot{z}_i^\mu a_\mu^b\left(z_i\right),\\
     a_\mu &\equiv & \Lambda^{-1} A_\mu .
\end{eqnarray*}

The solution to equation (\ref{ec5}) is
\begin{equation}\label{ITordenado}
    I_i^{a}(\tau)=\mathbf{T} \exp \left[-\Lambda \int_0^\tau d \tau^{\prime} R_i\left(\tau^{\prime}\right)\right] I_i^{a}(0).
\end{equation}
By replacing (\ref{ITordenado}) in (\ref{ec1}), and expanding the time-ordered exponential, we arrive at
\begin{equation}
    \begin{aligned}
2 \epsilon^{\mu \nu \rho} \partial_\nu a_\rho^a(x)= & -\Lambda \epsilon^{\mu \nu \rho} f^{a b c} a_\nu^b(x) a_\rho^c(x)+\sum_{i=1}^n \oint_{\gamma_i} d z^\mu \delta^3(x-z) I_i^a(0) \\
& -\Lambda \sum_{i=1}^n \oint_{\gamma_i} d z^\mu \int_0^z d z_1^{\mu_1} R_{\mu_1}^{a a_1}\left(z_1\right) \delta^3(x-z) I_i^{a_1}(0) \\
& +\Lambda^2 \sum_{i=1}^n \oint_{\gamma_i} d z^\mu \int_0^z d z_1^{\mu_1} \int_0^{z_1} d z_2^{\mu_2} R_{\mu_1}^{a a_1}\left(z_1\right) R_{\mu_2}^{a_1 a_2}\left(z_2\right)\\
&\delta^3(x-z) I_i^{a_2}(0) \\
& \vdots \\
& +(-\Lambda)^p \sum_{i=1}^n \oint_{\gamma_i} d z^\mu \int_0^z d z_1^{\mu_1} \ldots \int_0^{z_{p-1}} d z_p^{\mu_p} R_{\mu_1}^{a a_1}\left(z_1\right) \\
& \ldots R_{\mu_p}^{a_{p-1} a_p} \left(z_p\right) \delta^3(x-z) I_i^{a_p}(0) \\
& \vdots
\end{aligned}
\end{equation}
Next, if we replace $a_\mu^a$ by the power series
\begin{equation}
    a_\mu^a=\sum_{p=0}^{\infty} \Lambda^p a^{(p)a}_\mu,
\end{equation}
the $p$-th order term reads
\begin{equation}\label{pne0}
    \begin{aligned}
& 2 \epsilon^{\mu \nu \rho} \partial_\nu a_\rho^{(p)a}(x)=-\epsilon^{\mu \nu \rho} f^{a b c} \sum_{r, s=0}^{r+s=p-1} a_\nu^{(r)b} a_\rho^{(s)c}+ \\
& +\sum_{r=1}^p(-1)^r \sum_{i=1}^n \oint_{\gamma_i} d z^\mu \int_0^z d z_1^{\mu_1} \ldots \int_0^{z_{r-1}} d z_r^{\mu_r} \sum_{s_1, \ldots, s_r=o}^{s_1+\ldots+s_r=p-r} R_{\mu_1}^{\left(s_1\right){a a_1}}\left(z_1\right) \times \\
& \times R_{\mu_2}^{\left(s_2\right){a_1 a_2}}\left(z_2\right) \quad \ldots \quad R_{\mu_r}^{\left(s_r\right){a_{r-1} a_r}}\left(z_r\right) I_i^{a_r}(0) \delta^3(x-z), \\
&
\end{aligned}
\end{equation}
for $p \geq 1$. If $p=0$, the corresponding equation is
\begin{equation}\label{pe0}
    2 \epsilon^{\mu \nu \rho} \partial_\nu a_\rho^{(0)a}=\sum_{i=1}^n \oint_{\gamma_i} d z_i^\mu \delta^3\left(x-z_i\right) I_i^a(0) .
\end{equation}
At this point, a couple of comments are in order. First, note that in Eq. (\ref{pne0})
the right hand involves $a^{(q)}$ with $q<p$ so we can look for a recursive solution. Second, equations (\ref{pne0}) and (\ref{pe0}) have the same structure, namely, they have the form of Ampere's Law
\begin{equation}
    \epsilon^{\mu \nu \rho} \partial_\nu a_\rho^{(p)a}(x)=J^{(p){\mu a}}(x)
\end{equation}
so their solution is expressed as the Biot-Savart Law
\begin{equation}
    a^{(p)a}_\alpha (x)=-\frac{1}{4 \pi} \int d^3 x^{\prime} \epsilon_{\alpha \beta \gamma} J^{(p) \beta a}\left(x^{\prime}\right) \frac{\left(x-x^{\prime}\right)^\gamma}{\left|x-x^{\prime}\right|^3}+\partial_\alpha f^a(x),
\end{equation}
where $f^a(x)$ is arbitrary. \\

Now, as we have solved the equation of motion recursively, the on-shell action can be expressed as
\begin{equation}
    \begin{aligned}
S_{\text {OS}} & =S_{\text {CS,\ OS }}\\
& =\left.\frac{\Lambda}{2} \int d^3 x \epsilon^{\mu \nu \rho}\left(a_\mu^a \partial_\nu a_\rho^a+\frac{\Lambda}{3} f^{a b c} a_\mu^a a_\nu^b a_\rho^c\right)\right|_{\text {OS }} \\
& =\frac{\Lambda}{2} \sum_{p=0}^{\infty} S^{(p)} \Lambda^p,
\end{aligned}
\end{equation}
where
\begin{equation}\label{pesimo}
   \begin{aligned}
        & S^{(p)}=\int d^3 x \epsilon^{\mu \nu \rho}\left(\sum_{r, s}^{r+s=p}\left(a_\mu^{(r)a} \partial_\nu a_\rho^{(s)a}\right) + \frac{1}{3} f^{a b c} \sum_{r, s, q}^{r+s+q=p-1}\left(a_\mu^{(r)a} a_\nu^{(s)b} a_\rho^{(q)c} \right) \right).
   \end{aligned}
\end{equation}
It is worth noticing that Eq. (\ref{pesimo}), encodes the whole information about the link-invariants associated with the original topological action. More precisely, each order $p$ corresponds to an analytical expression of a LI which involves $p+2$ closed curves. For example, at $0^{th}$ order we have \cite{leal2002}
\begin{equation}
    S^{(0)}=\frac{1}{4} \sum_{i, j} I_i^a(0) I_j^a(0) L(i, j),
\end{equation}
where
\begin{equation}
    L(i, j) \equiv \frac{1}{4 \pi} \oint_{\gamma_i} d z^\mu \oint_{\gamma_j} d y^\rho \frac{(z-y)^\beta}{|z-y|^3} \epsilon_{\mu \nu \rho},
\end{equation}
is the GLN. Furthermore, at first order we obtain (see \cite{leal2002} for details)
\begin{equation}
    \begin{aligned}
S^{(1)}=-\frac{1}{4} \sum_{i, j, k} f^{a b c} I_i^a(0) I_j^b(0) I_k^c(0)\left\{\frac{1}{3} \int\right. & d^3 x \epsilon^{\mu \nu \rho} D_{i \mu}(x) D_{j \nu}(x) D_{k \rho}(x)+ \\
& \left.+\oint_{\gamma_i} d z^\mu \int_0^z d y^\nu D_{j \mu}(z) D_{k \nu}(y)\right\} ,
\end{aligned}
\end{equation}
where 
\begin{equation}\label{deriva}
    D_{i \alpha}(x) \equiv \frac{1}{4 \pi} \oint_{\gamma_i} d z^\gamma \frac{(x-z)^\beta}{|x-z|^3} \epsilon_{\alpha \beta \gamma} .
\end{equation}
The factor $f^{abc}I_i^a(0)I_j^b(0)I_k^c(0)$ in the above expression vanishes when the currents $I_i^a(0)$, $I_j^b(0)$, and $I_k^c(0)$ are linearly dependent and, as a consequence, $S^{(1)}$ is zero when the current consists of only one or two Wong particles. If we consider $SU(2)$ as gauge group and three Wong particles with orthonormal isovectors $I_i^a(0) = \delta_i^a$, the above expression can be written as
\begin{equation}
    \begin{aligned}
S^{(1)}(1,2,3)= & -\frac{1}{2} \int d^3 x \epsilon^{\mu \nu \rho} D_{1 \mu}(x) D_{2 \nu}(x) D_{3 \rho}(x)- \\
& -\frac{1}{2} \int d^3 x \int d^3 y\left\{T_1^{[\mu x, \nu y]} D_{2 \mu}(x) D_{3 \nu}(y)+\right.\\
& +T_2^{[\mu x, \nu y]} D_{3 \mu}(z) D_{1 \nu}(y)+ \\
& \left.+T_3^{[\mu x, \nu y]} D_{1 \mu}(z) D_{2 \nu}(y)\right\},
\end{aligned}
\label{mli}
\end{equation}
where we have introduced the bilocal object $T_{\gamma_i}^{\mu x, \nu y}$ associated with the curve $\gamma_i$
\begin{equation}\label{tobjeto1}
    T_{\gamma_i}^{\mu x, \nu y} \equiv \oint_{\gamma_i} d z^\mu \int_0^z d z^{\prime \nu} \delta^3(x-z) \delta^3\left(y-z^{\prime}\right),
\end{equation}
which satisfies
\begin{equation}
    T_{\gamma_i}^{[\mu x, \nu y]} \equiv \frac{1}{2}\left(T_{\gamma_i}^{\mu x, \nu y}-T_{\gamma_i}^{\nu x, \mu y}\right).
\end{equation}
It should be mentioned that (\ref{tobjeto1}) is one representative of an infinite family of $T$-objects which are known as coordinates for a extended loop space (or path coordinates), first introduced in \cite{gambini,gambini2}, used for a representation of quantum gravity. They are distributional objects of rank $n$ also known as path multi-tangents, It is worth mentioning that the $T$-object of one index is nothing more than the form factor or tangent distribution of a path. They are defined as
\begin{equation}
T_{i}^{\mu_1 x_1 \mu_2 x_2 \ldots \mu_n x_n} \equiv \oint_{\gamma_i} d z^{\mu_1} \int_0^z d z_1^{\mu_2} \int_0^{z_1} d z_2^{\mu_3} \ldots \int_0^{z_{n-2}} d z_{n-1}^{\mu_n} \delta^{(3)}\left(x_1-z\right) \delta^{(3)}\left(x_2-z_1\right) \delta^{(3)}\left(x_3-z_2\right) \ldots \delta^{(3)}\left(x_n-z_{n-1}\right),
\end{equation}
satisfying the differential constraint
\begin{equation}
    \frac{\partial}{\partial x_i^{\mu_i}} T_j^{\mu_1 x_1 \cdots \mu_i x_i \cdots \mu_n x_n}=\left(-\delta\left(x_i-x_{i-1}\right)+\delta\left(x_i-x_{i+1}\right)\right) T_j^{\mu_1 x_1 \cdots \mu_{i-1} x_{i-1} \mu_{i+1} x_{i+1} \cdots \mu_n x_n},
    \label{ld}
\end{equation}
where $x_0$ and $x_{n+1}$ correspond to the starting points of the closed path, and the Dirac deltas are defined in the same dimension as the manifold where the paths ``live''. Besides, they satisfies the algebraic constraint
\begin{equation}
    T_i^{\left\{\mu_1 \cdots \mu_k\right\} \mu_{k+1} \cdots \mu_n}=\sum_{P_k} T_i^{P_k\left(\mu_1 \cdots \mu_n\right)}=T_i^{\mu_1 \cdots \mu_k} T_i^{\mu_{k+1} \cdots \mu_n},
    \label{la}
\end{equation}
where the sum is taken over all permutations of the variables $\mu$ that preserve the ordering of $\mu_1, \cdots, \mu_k$ and $\mu_{k+1}, \cdots, \mu_n$ among themselves. \\

As demonstrated in \cite{leal2002}, Eq. (\ref{mli}) turns out to be, except for a factor, the Milnor's third linking number $\bar{\mu}(1,2,3)$, which detects (for example) the linking of the Borromeans'  rings. It is interesting to note that the third Milnor coefficient is defined whenever the three curves do not get linked (each other) in the Gauss sense, that is, $L(i,j)=0$ for any pair of considered curves. In fact, Milnor demonstrated that the third coefficient belongs to a sequence of infinite knot invariants $K_{n}$ that satisfy the restriction $K_{p}=0$ for $p<q$. This structure is repeated in our framework, as we will demonstrate later.
\\

To proceed with the perturbative analysis and obtain the LI associated with the on-shell second-order action $S^{(2)}$, it is necessary to introduce certain conventions that allow us to perform calculations more efficiently. Let us consider the object $g_{\mu x \text{ } \nu y}$, which is symmetric in its index pairs and is defined as
\begin{equation}
    \begin{aligned}
        g_{\mu x \text{ } \nu y} & \equiv-\frac{1}{4 \pi} \epsilon_{\mu \nu \rho} \frac{(x-y)^\rho}{|x-y|^3},
    \end{aligned}
    \label{g}
\end{equation}
which, together with $g^{\mu x \text{ } \nu y}$ defined by
\begin{equation}
    \begin{aligned}
g^{\mu x \text{ } \nu y} & \equiv \epsilon^{\mu \nu \rho} \partial_\rho \delta(x-y), \\
\end{aligned}
\end{equation}
naturally appear in the solution of the differential constraint obeyed by the $T$-objects and constitute a metric in the space of transverse rank-one vector densities \cite{gambini2}. Furthermore,
let us denote the dependence of a tensorial function on a continuous variable by placing an index indicating that variable, namely
\begin{equation}
    A_{\mu \nu \ldots \rho}(x, y, \ldots, z) \equiv A_{\mu x \text{ } \nu y \text{ } \ldots \text{ } \rho z}.
\end{equation}
Now it is possible to establish a kind of generalized Einstein convention (that integrates repeated continuous variables instead of summing them) as
\begin{equation}
    A_{\mu x} B^{\mu x \text{ } \nu y \ldots} \equiv \sum_\mu \int A_{\mu x} B^{\mu x \text{ } \nu y \cdots} d^3 x=\sum_\mu \int A_\mu(x) B^{
    \mu \nu \cdots}(x, y \ldots) d^3 x.
\end{equation}
Now, let us establish that when repeated indices that are not integrated, we will place a ``bar'' above such a letter in the following way
\begin{equation}
    A_{\mu x \text{ } \nu \bar{y}} B^{\mu x \text{ } \nu \bar{y}} \equiv \sum_\mu \int A_{\mu x \text{ } \nu \bar{y}} B^{\mu x \text{ } \nu \bar{y}} d^3 x.
\end{equation}
Finally, let us indicate with a lower case latin letter the set of discrete-continuous variables, namely
\begin{equation}
    A^{\mu x \text{ } \nu y \text{ } \ldots \text{ } \rho z} \equiv A^{a b \ldots c}.
\end{equation}

Note that these conventions are useful and allow us to write known quantities in a compact way.  For example, the quantity in (\ref{deriva}) can be written as
\begin{equation}
    D_\mu(x, \gamma) \equiv D_{i \text{ } \mu x}=-g_{\mu x \text{ } \nu y} T_i^{\nu y}.
    \label{od}
\end{equation}
Similarly, the GLN and the Biot-Savart law of the non-Abelian theory take the form
\begin{eqnarray}
L(i, j) &=&\frac{1}{4 \pi} \oint_{\gamma_i} d z^\mu \oint_{\gamma_j} d y^\rho \frac{(z-y)^\nu}{|z-y|^3} \epsilon_{\mu \nu \rho} = T^{\mu x}_i g_{\mu x \text{ } \nu y} T^{\nu y}_j \label{gln}\\
a_\mu^{(p)a}(x) &=&\frac{1}{4 \pi} \int d^3 x^{\prime} \epsilon_{\mu \nu \rho} J^{(p) \nu a}\left(x^{\prime}\right) \frac{\left(x-x^{\prime}\right)^\rho}{\left|x-x^{\prime}\right|^3} =-J^{(p) \nu y} g_{\mu x \text{ } \nu y}. \label{eompg}
\end{eqnarray}

With all these tools at hand, and using the equation (\ref{pesimo}) for $p=2$, we proceed to rewrite the second-order on-shell action in a contracted form as follows
\begin{equation}
    S^{(2)}=\varepsilon^{\mu \nu \rho}\left[2 \partial_\nu \vec{a}_{\rho x}^{(2)} \cdot \vec{a}_{\mu x}^{(0)}+\partial_\nu \vec{a}_{\rho x}^{(1)} \cdot \vec{a}_{\mu x}^{(1)}+\vec{a}_{\mu x}^{(1)} \cdot\left(\vec{a}_{\nu x}^{(0)} \times \vec{a}_{\rho x}^{(0)}\right)\right].
    \label{sos2o}
\end{equation}
In what follows, we will consider the simplest nontrivial case, which consists of restricting the gauge group to $SU(2)$. Besides, we will denote with an arrow any quantity that has components in the internal space. For example, the isovectors $I^{a}_{i}$ will be indicated as $\vec{I}_{i}$.  Therefore, by using the equation (\ref{eompg}) for $p=0,1$ and substituting into (\ref{sos2o}), we can express $S^{(2)}$ solely in terms of the path coordinates and the metric as
\begin{equation}
    \begin{aligned}
S^{(2)} & =\frac{1}{8} \sum_{i, j, k, l}\left(\left[\vec{I}_i \times\left(\vec{I}_j \times \vec{I}_k\right)\right] \cdot \vec{I}_l\right) g_{\mu x \text{ } \alpha y} \\
& \left\{\frac{3}{8} \varepsilon^{\mu \nu \rho} \varepsilon^{\alpha \beta \gamma} g_{\nu x \text{ } \mu_1 x_1} g_{\rho x \text{ } \mu_4 x_4} g_{\beta y \text{ } \mu_2 x_2} g_{\gamma y \text{ } \mu_3 x_3} T_i^{\mu_1 x_1} T_j^{\mu_2 x_2} T_k^{\mu_3 x_3} T_l^{\mu_4 x_4}\right. \\
& +\varepsilon^{\mu \nu \rho}\left(T_j^{\alpha y \text{ }  \mu_1 x_1}-\frac{1}{2} T_j^{\mu_1 x_1 \text{ } \alpha y}\right) g_{\nu x \text{ } \mu_2 x_2} g_{\rho x \text{ } \mu_4 x_4} g_{\mu_1 x_1 \text{ } \mu_3 x_3} T_i^{\mu_2 x_2} T_k^{\mu_3 x_3} T_l^{\mu_4 x_4} \\
& -T_j^{\alpha y \text{ } \mu_2 x_2}\left(T_i^{\mu_1 x_1 \text{ } \mu x}-\frac{1}{2} T_i^{\mu x \text{ } \mu_1 x_1}\right) g_{\mu_2 x_2 \text{ } \mu_3 x_3} g_{\mu_1 x_1 \text{ } \mu_4 x_4} T_k^{\mu_3 x_3} T_l^{\mu_4 x_4} \\
& +T_j^{\mu x \text{ } \mu_1 x_1 \text{ } \mu_2 x_2} g_{\mu_1 x_1 \text{ } \mu_3 x_3} g_{\mu_2 x_2 \text{ } \mu_4 x_4} T_i^{\mu_3 x_3} T_k^{\mu_4 x_4} T_l^{\alpha y}\biggr\}.
\end{aligned}
\end{equation}
Finally, decomposing the two-index $T$-objects into their symmetric and antisymmetric parts $\left(T_i^{ab}=T_i^{(ab)}+T_i^{[ab]}\right)$, and using the fact that the symmetric part factorizes into two 1-index $T$-object $\left(T_i^{(ab)}=T_i^a T_i^b\right)$ according to the algebraic constraint (\ref{la}) we obtain
\begin{equation}\label{segundo-or}
    \begin{split}
        S^{(2)} = \enskip \frac{1}{8} \sum_{i, j, k, l}\left(\left(\vec{I}_i(0) \times\left(\vec{I}_j(0) \times \vec{I}_k(0)\right)\right) \cdot \vec{I}_l(0)\right) & \left\{\frac{3}{8} \varepsilon^{\mu \nu \rho} \varepsilon^{\alpha \beta \gamma}g_{\nu x \text{ }\mu_1 x_1} g_{\beta y \text{ } \mu_2 x_2} g_{\gamma y \text{ } \mu_3 x_3} g_{\rho x \text{ } \mu_4 x_4} T_i^{\mu_1 x_1} T_j^{\mu_2 x_2} T_k^{\mu_3 x_3} T_l^{\mu_4 x_4}\right.\\
        &+\frac{3}{2} \varepsilon^{\mu \nu \rho} T_j^{\left[\alpha y \text{ }\mu_1 x_1\right]} g_{\nu x \text{ }\mu_2 x_2} g_{\mu_1 x_1 \text{ } \mu_3 x_3} g_{\rho x \text{ } \mu_4 x_4} T^{\mu_2 x_2}_i T^{\mu_3 x_3}_k T^{\mu_4 x_4}_l\\
        &+\frac{3}{2} T_i^{[\mu x \text{ } \mu_1 x_1]} T_j^{\left[\alpha y \text{ } \mu_2 x_2\right]}  g_{\mu_2 x_2 \text{ } \mu_3 x_3} g_{\mu_1 x_1 \text{ } \mu_4 x_4} T^{\mu_3 x_3}_k T^{\mu_4 x_4}_l\\
        &+ T_j^{\mu x \text{ } \mu_1 x_1 \text{ } \mu_2 x_2} g_{\mu_1 x_1 \text{ } \mu_3 x_3} g_{\mu_2 x_2 \text{ } \mu_4 x_4} T^{\mu_3 x_3}_i T^{\mu_4 x_4}_k T^{\alpha y}_l \biggr \} g_{\mu x \text{ } \alpha y},
    \end{split}
\end{equation}
which by using (\ref{od}) can be written as
\begin{equation}
     \begin{split}
         S^{(2)} = \enskip \frac{1}{8} \sum_{i, j, k, l}\left(\left(\vec{I}_i \times\left(\vec{I}_j \times \vec{I}_k\right)\right) \cdot \vec{I}_l\right) & \left\{\frac{3}{8} \varepsilon^{\mu \nu \rho} \varepsilon^{\alpha \beta \gamma} g_{\mu x \text{ } \alpha y} D_{i \text{ } \nu x} D_{j  \text{ } \beta y} D_{k \text{ } \gamma y} D_{l \text{ } \rho x} \right.\\
         &-\frac{3}{2} \varepsilon^{\mu \nu \rho} g_{\mu x \text{ } \alpha y} T_j^{\left[\alpha y \text{ }\mu_1 x_1\right]}  D_{i  \text{ } \nu x} D_{k \text{ } \mu_1 x_1} D_{l \text{ } \rho x} \\
         &+\frac{3}{2} g_{\mu x \text{ } \alpha y} T_i^{[\mu x \text{ } \mu_1 x_1]} T_j^{\left[\alpha y \text{ } \mu_2 x_2\right]}   D_{k \text{ } \mu_2 x_2} D_{l \text{ } \mu_1 x_1}\\
         &- T_j^{\mu x \text{ } \mu_1 x_1 \text{ } \mu_2 x_2} D_{i \text{ } \mu_1 x_1} D_{k \text{ } \mu_2 x_2} D_{l \text{ } \mu x} \biggr \}.
     \end{split}
    \label{aosna2}
\end{equation}

Now, we will demonstrate that the invariant (\ref{aosna2}) is well-defined whenever both the GLN and the third Milnor coefficient vanish which implies the consistency of the theory in terms of its integrability conditions that maintains gauge invariance. It is worth noticing that, due to the proposed perturbative method we will have an equation of motion and an integrability condition for each order $p$ and we will relate this fact to the definition of Milnor's higher order linking coefficients \cite{milnor}. \\

The equation of motion (\ref{pe0}) can be written as
\begin{equation}
    \varepsilon^{\mu \nu \rho} \partial_\nu \vec{a}^{(0)}_{\rho x} =   \frac{1}{2} \sum_{i=1}^n T_i^{\mu x} \vec{I}_i(0),
    \label{eom0}
\end{equation}
so taking the divergence yields
\begin{equation}
    \partial_\mu T_i^{\mu x} = 0,
    \label{cons1}
\end{equation}
which means that the curves must be closed. Now, if we use the equation (\ref{pne0}) we can write the first-order equation of motion as 
\begin{equation}
    \varepsilon^{\mu \nu \rho} \partial_\nu \vec{a}^{(1)}_{\rho x} = -\frac{1}{2} \varepsilon^{\mu \nu \rho}\left(\vec{a}^{(0)}_{\nu \bar{x}} \times \vec{a}^{(0)}_{\rho \bar{x}}\right)-\frac{1}{2} \sum_{i=1}^n T_i^{\mu x \text{ } \mu_1 x_1} \left(\vec{a}^{(0)}_{\mu_1 x_1} \times \vec{I}_i\right),
    \label{eompo1}
\end{equation}
so taking the divergence entails
\begin{equation}
    0 = \vec{a}^{(0)}_{\mu \bar{x}} \times \left(\varepsilon^{\mu \nu \rho} \partial_\nu \vec{a}^{(0)}_{\rho \bar{x}} \right) - \frac{1}{2} \sum_{i=1}^n \partial_\mu T_i^{\mu x \text{ } \mu_1 x_1} \left(\vec{a}^{(0)}_{\mu_1 x_1} \times \vec{I}_i\right).
    \label{deompo}
\end{equation}
From the general differential constraint (\ref{ld}), we can write
\begin{equation}
    \partial_\mu T_i^{\mu x \text{ } \mu_1 x_1} =\left(-\delta^3\left(x-x_i\right)+\delta^3\left(x-x_1\right)\right) T_i^{\mu_1 x_1}
    \label{ld2},
\end{equation}
and as a consequence (\ref{deompo}) can be expressed as
\begin{equation}
    \frac{1}{2} \sum_{j=1}^n \delta^3\left(x-x_j\right) T_j^{\mu_1 x_1} \left(\vec{a}^{(0)}_{\mu_1 x_1} \times \vec{I}_j\right) = 0.
    \label{dompor}
\end{equation}
Finally, by combining  (\ref{gln}), and (\ref{dompor}) we arrive at
\begin{equation}
    \frac{1}{4} \sum_{i,j} \delta^3\left(x-x_j\right) (I_i \times I_j) L(i,j) = 0.
\end{equation}
which requires
\begin{equation}\label{consg}
    L(i,j) = 0,
\end{equation}
namely, the GLN must vanish as required.\\

Similarly, using the equation (\ref{pne0}) we can write the second-order equation of motion as 
\begin{equation}
    2 \varepsilon^{\mu \nu \rho} \partial_\nu \vec{a}^{(2)}_{\rho x}= -2 \varepsilon^{\mu \nu \rho} \vec{a}^{(1)}_{\nu {\bar{x}}} \times \vec{a}^{(0)}_{\rho \bar{x}}  -\sum_{i=1}^n T_i^{\mu x \text{ } \mu_1 x_1} \vec{a}^{(1)}_{\mu_1 x_1} \times \vec{I}_i + \sum_{i=1}^n T_i^{\mu x \text{ } \mu_1 x_1 \text{ } \mu_2 x_2} \left[\vec{a}^{(0)}_{\mu_1 x_1} \times\left(\vec{a}^{(0)}_{\mu_2 x_2} \times \vec{I}_i\right)\right],
\end{equation}
Now, if we take its divergence and use the differential constraint (\ref{ld}), the equation of motion (\ref{eompo1}), and the following vector identities
$$ \epsilon^{\mu \nu \rho}\left(\vec{a}_{\mu \bar{x}}^{(0)} \times \vec{a}_{\nu \bar{x}}^{(0)}\right) \times \vec{a}_{\rho \bar{x}}^{(0)} = 0, $$
$$ \left(\vec{a}_{\mu \bar{x}}^{(0)} \times \vec{a}_{\nu \bar{x}}^{(0)}\right) \times \vec{a}_{\rho \bar{x}}^{(0)}=\vec{a}_{\nu \bar{x}}^{(0)}\left(\vec{a}_{\mu \bar{x}}^{(0)} \cdot \vec{a}_{\rho \bar{x}}^{(0)}\right)-\vec{a}_{\mu \bar{x}}^{(0)}\left(\vec{a}_{\nu \bar{x}}^{(0)} \cdot \vec{a}_{\rho \bar{x}}^{(0)}\right), $$
we arrive at
\begin{equation}
    T_i^{\mu_1 x_1} \vec{a}_{\mu_1 x_1}^{(1)} \times \vec{I}_i-T_i^{\mu_1 x_1 \text{ } \mu_2 x_2} \vec{a}_{\mu_1 x_1}^{(0)} \times\left(\vec{a}_{\mu_2 x_2}^{(0)} \times \vec{I}_i\right) = 0.
    \label{cons3}
\end{equation}
Finally, by replacing (\ref{eom0}) and the solution to the equation of motion (\ref{eompo1}) into (\ref{cons3}), we obtain
\begin{equation}
    \begin{aligned}
    & \sum_{j, k}\left[\left(\vec{I}_j \times \vec{I}_k\right) \times \vec{I}_i\right]\left\{\frac{1}{2} \epsilon^{\mu \nu \rho} T_i^a T_j^b T_k^c g_{\mu x \text{ } a} g_{\nu x \text{ } b} g_{\rho x \text{ } c}+T_i^a T_j^{c d} T_k^b g_{c a} g_{d b}\right\} \\
    & -\sum_{j, k}\left[\vec{I}_j \times\left(\vec{I}_k \times \vec{I}_i\right)\right] T_j^a T_i^{c d} T_k^b g_{c a} g_{d b}=0 .
\end{aligned}
\label{cons4}
\end{equation}
Note that, if we take the dot product between the equation (\ref{cons4}) and the vector $\vec{I}_i$, and using the following vector identities
$$\left[\left(\vec{I}_j \times \vec{I}_k\right) \times \vec{I}_i\right] \cdot \vec{I}_i=0,$$
$$\left[\vec{I}_j \times\left(\vec{I}_k \times \vec{I}_i\right)\right] \cdot \vec{I}_i=\left(\vec{I}_i \cdot \vec{I}_j\right)\left(\vec{I}_i \cdot \vec{I}_k\right)-\left|\vec{I}_i\right|^2\left(\vec{I}_j \cdot \vec{I}_k\right),$$
the consistency condition turns out to be
\begin{equation}
    \sum_{j, k}\left[\left(\vec{I}_i \cdot \vec{I}_j\right)\left(\overrightarrow{I_i} \cdot \vec{I}_k\right)-\left|\vec{I}_i\right|^2\left(\vec{I}_j \cdot \vec{I}_k\right)\right] T_j^a T_i^{c d} T_k^b g_{c a} g_{d b}=0.
    \label{cons5}
\end{equation}
Now, as the expression is symmetric between the indices $j,k$, Eq. (\ref{cons5}) can be written, after symmetrization and by using  $T^{(ab)}= T^a T^b$, as
\begin{equation}
    \sum_{j, k} T_j^a T_i^c T_i^d T_k^b g_{c a} g_{d b}=\sum_{j, k}\left(T_j^a g_{c a} T_i^c\right)\left(T_i^d g_{d b} T_k^b\right)=\sum_{j, k} L(i, j) L(i, k)=0,
\end{equation}
so $L(i,j)=0$ as required. In contrast, if we take the vector product between Eq. (\ref{cons4}) and $\vec{I}_i$,  and consider the following vector identities
$$ \left[\left(\vec{I}_j \times \vec{I}_k\right) \times \vec{I}_i\right] \times \vec{I}_i=\left(\vec{I}_k \times \vec{I}_i\right)\left(\vec{I}_i \cdot \vec{I}_j\right)-\left(\vec{I}_j \times \vec{I}_i\right)\left(\vec{I}_i \cdot \vec{I}_k\right), $$
$$\left[\vec{I}_j \times\left(\vec{I}_k \times \vec{I}_i\right)\right] \times \vec{I}_i=\left(\vec{I}_k \times \vec{I}_i\right)\left(\vec{I}_i \cdot \vec{I}_j\right),$$
the consistency condition is now given by
\begin{equation}
    \begin{aligned}
    & \sum_{i, j, k}\left[\left(\vec{I}_k \times \vec{I}_i\right)\left(\overrightarrow{I_i} \cdot \overrightarrow{I_j}\right)\right]\left\{\frac{1}{2} \epsilon^{\mu \nu \rho} T_i^a T_j^b T_k^c g_{\mu x a} g_{\nu x b} g_{\rho x c}+\left(T_i^a T_j^{[c d]} T_k^b-T_j^a T_i^{[c d]} T_k^b\right) g_{c a} g_{d b}\right\} \\
    & -\sum_{i, j, k}\left[\left(\vec{I}_j \times \vec{I}_i\right)\left(\vec{I}_i \cdot \overrightarrow{I_k}\right)\right]\left\{\frac{1}{2} \epsilon^{\mu \nu \rho} T_i^a T_j^b T_k^c g_{\mu x a} g_{\nu x b} g_{\rho x c}+T_i^a T_j^{[cd]} T_k^b g_{c a} g_{d b}\right\}=0,
    \end{aligned}
\end{equation}
where we have appropriately antisymmetrized the term within braces, taking into account the antisymmetry present in the coefficients of the iso-charges ($(i, k)$ in the first sum and $(i, j)$ in the second). Interchanging the indices $(j \leftrightarrow k)$ in the second term of the last equation, and using (\ref{od}) we obtain
\begin{equation}
    \sum_{i, j, k} 2f_{i,j,k} \bar{\mu}(i,j,k)  = 0,
    \label{cons6}
\end{equation}
where $f_{i,j,k}=\left(\vec{I}_k \times \vec{I}_i\right)\left(\vec{I}_i \cdot \vec{I}_j\right)$ and
\begin{equation}
\bar{\mu}(i,j,k)= - \frac{1}{2}\left[ \epsilon^{\mu \nu \rho} D_{i \text{ } \mu x} D_{j \text{ } \nu x} D_{k \text{ } \rho x} + \left( T_i^{[\mu x \text{ } \nu y]} D_{j \text{ } \mu x} D_{k \text{ } \nu y} +  T_j^{[\mu x \text{ } \nu y]} D_{i \text{ } \nu y} D_{k \text{ } \mu x} + T_k^{[\mu x \text{ } \nu y]} D_{i \text{ } \mu x} D_{j \text{ } \nu y} \right) \right].
\end{equation}
Note that as the function of the iso-currents $f_{i,j,k}$ in the equation (\ref{cons6}) is not generally zero, it implies that the third Milnor's coefficient, $\bar{\mu}(i,j,k)$, vanishes as required. Therefore, $S^{(2)}$ corresponds to the LI
of four closed loops linked in space-time (as we will show later), but not in the way of the GLN nor the third Milnor's coefficient. In this regard, if we are intended to relate the second order contribution to some Milnor Coefficient then results (\ref{consg}) and (\ref{cons6}) present great arguments forward to define this invariant.\\

Before addressing the geometrical interpretation of (\ref{segundo-or}), we would like to tackle the problem of the Abelian intermediate action that reproduces the same invariant we have just found. This treatment is similar to what has already been developed in \cite{leal-intermedia}, where it is constructed an Abelian model that is able to reproduce the first-order action associated with the third Milnor's coefficient.

\section{Intermediate second--order action }\label{ia}
As Abelian intermediate action, we propose
\begin{equation}
   \begin{split}
       S = \enskip & -6 \int d^3x \enskip \vec{\Lambda}_{\mu x} \cdot \left[\varepsilon^{\mu \nu \rho} \partial_\nu \vec{\lambda}_{\rho x}-\frac{1}{2} \sum_i^n T_i^{\mu x} \vec{I}_i\right]\\
       & + \frac{3}{4} \int d^3x \int d^3 y \enskip \varepsilon^{\mu \nu \rho} \varepsilon^{\alpha \beta \gamma} g_{\mu x \text{ }\alpha y}\left[\vec{\lambda}_{\nu x} \times \left(\vec{\lambda}_{\beta y} \times \vec{\lambda}_{\gamma y}\right)\right] \cdot \vec{\lambda}_{\rho x} \\
       & + \frac{3}{2} \sum_{i=1}^n \int d^3x \int d^3 y \int d^3 x_1 \enskip \varepsilon^{\mu \nu \rho} g_{\mu x \text{ } \alpha y} T_i^{\left[\alpha y \text{ } \mu_1 x_1\right]} \left[\vec{\lambda}_{\nu x} \times\left(\vec{\lambda}_{\mu_1 x_1} \times \vec{I}_i\right)\right] \cdot \vec{\lambda}_{\rho x}\\
       & +\frac{3}{4} \sum_{i, j} \int d^3x \int d^3 y \int d^3 x_1 \int d^3 x_2 \enskip g_{\mu x \text{ } \alpha y} T_i^{\left[\mu_1 x_1 \text{ } \mu x\right]} T_j^{\left[\alpha y \text{ } \mu_2 x_2\right]} \left[\vec{I}_i \times\left(\vec{\lambda}_{\mu_2 x_2} \times \vec{I}_j\right)\right] \cdot \vec{\lambda}_{\mu_1 x_1}\\
       & + \sum_{i=1}^n \int d^3x  \int d^3 x_1 \int d^3 x_2 \enskip T_i^{\mu x \text{ } \mu_1 x_1 \text{ } \mu_2 x_2}\left[\vec{\lambda}_{\mu_1 x_1} \times \left(\vec{\lambda}_{\mu_2 x_2} \times \vec{I}_i\right)\right] \cdot \vec{\lambda}_{\mu x},
   \end{split}
   \label{aim}
\end{equation}
where $\vec{\lambda}_{\mu x}$ and $\vec{\Lambda}_{\mu x}$ are two independent sets of Abelian gauge fields, labeled by latin letters running from 1 to 3 (here we are using vector notation for internal indices, i.e., the fields can be seen as $\lambda^a_{\mu x}$ and $\Lambda^a_{\mu x}$), and the current $\vec{I}_i$, corresponding to the $i$-th particle, is a vector which belongs to the internal space.\\

By using the notation for discrete-continuous indices introduced previously, the intermediate action reads
\begin{equation}
   \begin{split}
       S = \enskip & -6 \vec{\Lambda}_{\mu x} \cdot \left[\varepsilon^{\mu \nu \rho} \partial_\nu \vec{\lambda}_{\rho x}-\frac{1}{2} \sum_i^n T_i^{\mu x} \vec{I}_i\right]\\
       & + \frac{3}{4} \varepsilon^{\mu \nu \rho} \varepsilon^{\alpha \beta \gamma} g_{\mu x \text{ }\alpha y}\left[\vec{\lambda}_{\nu x} \times \left(\vec{\lambda}_{\beta y} \times \vec{\lambda}_{\gamma y}\right)\right] \cdot \vec{\lambda}_{\rho x} \\
       & + \frac{3}{2} \sum_{i=1}^n \varepsilon^{\mu \nu \rho} g_{\mu x \text{ } \alpha y} T_i^{\left[\alpha y \text{ } \mu_1 x_1\right]} \left[\vec{\lambda}_{\nu x} \times\left(\vec{\lambda}_{\mu_1 x_1} \times \vec{I}_i\right)\right] \cdot \vec{\lambda}_{\rho x}\\
       & +\frac{3}{4} \sum_{i, j} g_{\mu x \text{ } \alpha y} T_i^{\left[\mu_1 x_1 \text{ } \mu x\right]} T_j^{\left[\alpha y \text{ } \mu_2 x_2\right]} \left[\vec{I}_i \times\left(\vec{\lambda}_{\mu_2 x_2} \times \vec{I}_j\right)\right] \cdot \vec{\lambda}_{\mu_1 x_1}\\
       & + \sum_{i=1}^n T_i^{\mu x \text{ } \mu_1 x_1 \text{ } \mu_2 x_2}\left[\vec{\lambda}_{\mu_1 x_1} \times \left(\vec{\lambda}_{\mu_2 x_2} \times \vec{I}_i\right)\right] \cdot \vec{\lambda}_{\mu x}.
   \end{split}
\end{equation}

After variations of the action  with respect to the field $\vec{\Lambda}_{\mu x}$,  we obtain
\begin{equation}
    \varepsilon^{\mu \nu \rho} \partial_\nu \vec{\lambda}_{\rho x} = \frac{1}{2} \sum_i^n T_i^{\mu x} \vec{I}_i,
    \label{eom1}
\end{equation}
whose solution is given by
\begin{equation}
    \vec{\lambda}_{\mu x}=\frac{1}{2} \sum_i D_{i \text{ } \mu x} \vec{I}_i= -\frac{1}{2} \sum_i g_{\mu x \text{ } \nu y} T_i^{\nu y} \vec{I}_i = -\frac{1}{2} \sum_i \frac{1}{4 \pi} \oint_{\gamma_i} d z^\gamma \frac{(x-z)^\beta}{|x-z|^3} \epsilon_{\mu \beta \gamma} \vec{I}_i.
    \label{seom1}
\end{equation}
Now, variations with respect to $\vec{\lambda}$ lead to
\begin{equation}
    \begin{split}
        2 \varepsilon^{\mu \nu \rho} \partial_\nu \vec{\Lambda}_{\rho x} = &- \varepsilon^{\mu \nu \rho} \varepsilon^{\alpha \beta \gamma} g_{\rho \bar{x} \text{ }\alpha y}\left[\vec{\lambda}_{\nu \bar{x}} \times\left(\vec{\lambda}_{\beta y} \times \vec{\lambda}_{\gamma y}\right)\right] \\
        &- \sum_{i=1}^n \varepsilon^{\mu \nu \rho} g_{\rho \bar{x} \text{ } \alpha y} T_i^{\left[\alpha y \text{ } \mu_1 x_1\right]} \left[\vec{\lambda}_{\nu \bar{x}} \times\left(\vec{\lambda}_{\mu_1 x_1} \times \vec{I}_i\right)\right] \\
        &- \frac{1}{2} \sum_{i=1}^n \varepsilon^{\alpha \beta \gamma} g_{\mu_1 x_1 \text{ } \alpha y} T_i^{\left[\mu x \text{ } \mu_1 x_1 \right]} \left[\left(\vec{\lambda}_{\beta y} \times \vec{\lambda}_{\gamma y}\right) \times \vec{I}_i\right] \\
        &- \frac{1}{2} \sum_{i, j} g_{\mu_1 x_1 \text{ } \alpha y} T_i^{\left[\mu x \text{ } \mu_1 x_1\right]} T_j^{\left[\alpha y \text{ } \mu_2 x_2\right]} \left[\left(\vec{\lambda}_{\mu_2 x_2} \times{\vec{I}_j}\right) \times{\vec{I}_i}\right] \\
        &+ \sum_{i=1}^n T_i^{\langle [\mu x \text{ } \mu_1 x_1] \text{ } \mu_2 x_2 \rangle} \left[\vec{\lambda}_{\mu_1 x_1} \times\left(\vec{\lambda}_{\mu_2 x_2} \times \vec{I}_i\right)\right],
    \end{split}
    \label{eom2}
\end{equation}

where
\begin{eqnarray}
     T_i^{\langle [\mu x \text{ } \mu_1 x_1] \text{ } \mu_2 x_2 \rangle} &=& \frac{2}{3} \left( T_i^{[\mu x \text{ } \mu_1 x_1] \text{ } \mu_2 x_2} - T_i^{[\mu_1 x_1 \text{ }\mu_2 x_2] \text{ }\mu x} \right)\\
    T_i^{[\mu x \text{ } \mu_1 x_1] \text{ } \mu_2 x_2} &=& \frac{1}{2} \left( T_i^{\mu x \text{ } \mu_1 x_1 \text{ } \mu_2 x_2} - T_i^{\mu_1 x_1 \text{ } \mu x \text{ } \mu_2 x_2} \right).
    \label{taci}
\end{eqnarray}

Note that, equation (\ref{eom2}) can be rewritten as
\begin{equation}
    \begin{split}
        2 \varepsilon^{\mu \nu \rho} \partial_\nu \vec{\Lambda}_{\rho x}= &-2 \varepsilon^{\mu \nu \rho} \vec{a}_{\nu {\bar{x}}} \times \vec{\lambda}_{\rho \bar{x}}  -\sum_{i=1}^n T_i^{[\mu x \text{ } \mu_1 x_1]} \vec{a}_{\mu_1 x_1} \times \vec{I}_i\\
        &+ \sum_{i=1}^n T_i^{\langle [\mu x \text{ } \mu_1 x_1] \text{ } \mu_2 x_2 \rangle} \left[\vec{\lambda}_{\mu_1 x_1} \times\left(\vec{\lambda}_{\mu_2 x_2} \times \vec{I}_i\right)\right],
    \end{split}
\end{equation}
where we have defined
\begin{equation}
    \vec{a}_{\mu x} = \frac{1}{2} \varepsilon^{\alpha \beta \gamma}\left(\vec{\lambda}_{\beta y} \times \vec{\lambda}_{\gamma y}\right) g_{\mu x \text{ } \alpha y}+\frac{1}{2} \sum_{i=1}^n T_i^{[\alpha y \text{ } \mu_1 x_1]}\left(\vec{\lambda}_{\mu_1 x_1} \times \vec{I}_i\right) g_{\mu x \text{ } \alpha y}.
    \label{eom11}
\end{equation}
Now, contracting with the inverse metric tensor and setting $I_{lk}=\delta_{lk}$ we arrive at
\begin{equation}\label{eompo}
    \varepsilon^{\mu \nu \rho} \partial_\nu a_{i \rho}(x) = -\frac{1}{2} \varepsilon^{\mu \nu \rho} \varepsilon^{ijk} \lambda_{j \nu}(x) \lambda_{k \rho}(x) +\frac{1}{2} \int d^3y \varepsilon^{ijk} T_j^{[\mu x \text{ } \nu y]} \lambda_{k \nu}(y),
    \
\end{equation}
which is exactly the equation of motion obtained from  (\ref{pne0}) with $p=1$. Finally, replacing (\ref{seom1}) and (\ref{eom2}) into (\ref{aim}), the on-shell action is written as
\begin{equation}\label{on-shell-in}
    \begin{split}
        S_{On-Shell} = \enskip \frac{1}{8} \sum_{i, j, k, l}\left(\left(\vec{I}_i \times\left(\vec{I}_j \times \vec{I}_k\right)\right) \cdot \vec{I}_l\right) & \left\{\frac{3}{8} \varepsilon^{\mu \nu \rho} \varepsilon^{\alpha \beta \gamma}g_{\nu x \text{ }\mu_1 x_1} g_{\beta y \text{ } \mu_2 x_2} g_{\gamma y \text{ } \mu_3 x_3} g_{\rho x \text{ } \mu_4 x_4} T_i^{\mu_1 x_1} T_j^{\mu_2 x_2} T_k^{\mu_3 x_3} T_l^{\mu_4 x_4}\right.\\
        &+\frac{3}{2} \varepsilon^{\mu \nu \rho} T_j^{\left[\alpha y \text{ }\mu_1 x_1\right]} g_{\nu x \text{ }\mu_2 x_2} g_{\mu_1 x_1 \text{ } \mu_3 x_3} g_{\rho x \text{ } \mu_4 x_4} T^{\mu_2 x_2}_i T^{\mu_3 x_3}_k T^{\mu_4 x_4}_l\\
        &+\frac{3}{2} T_i^{[\mu x \text{ } \mu_1 x_1]} T_j^{\left[\alpha y \text{ } \mu_2 x_2\right]}  g_{\mu_2 x_2 \text{ } \mu_3 x_3} g_{\mu_1 x_1 \text{ } \mu_4 x_4} T^{\mu_3 x_3}_k T^{\mu_4 x_4}_l\\
        &+ T_j^{\mu x \text{ } \mu_1 x_1 \text{ } \mu_2 x_2} g_{\mu_1 x_1 \text{ } \mu_3 x_3} g_{\mu_2 x_2 \text{ } \mu_4 x_4} T^{\mu_3 x_3}_i T^{\mu_4 x_4}_k T^{\alpha y}_l \biggr \} g_{\mu x \text{ } \alpha y}.
    \end{split}
\end{equation}
which coincides with (\ref{segundo-or}), as expected.\\

The consistency of (\ref{on-shell-in}) is ensured as both the GLN and the third Milnor's coefficient vanish as we demonstrate in what follows. By taking the divergence of equation (\ref{eom1}), we arrive at 
\begin{equation}
    \partial_\mu T_i^{\mu x} = 0,
\end{equation}
which coincides with the differential constraint of the equation (\ref{ld}), which demonstrates the consistency of the equation (\ref{eom1}). This reflects the gauge invariance of the action (\ref{aim}) under gauge transformations.
\begin{equation}
    \Lambda_\mu^i (x) \enskip \longrightarrow \enskip \Lambda_\mu^i (x) + \partial_\mu \Omega^i(x),
    \label{gt1}
\end{equation}
where each $\Omega^i(x)$ is a scalar function. Similarly, by taking the divergence of (\ref{eompo})
\begin{equation}
    \frac{1}{4} \sum_{i,j} \delta^3\left(x-x_j\right) (I_i \times I_j) L(i,j) = 0.
    \label{ccsoa}
\end{equation}
where $L(i,j)$ is the Gauss linking number which must vanish as expected. Then, taking the divergence of (\ref{eom2}) and using (\ref{ccsoa}), we arrive at
\begin{equation}\label{milnoAbe}
    \sum_{i, j, k} 2f_{i,j,k} \bar{\mu}(i,j,k)  = 0,
\end{equation}
where $f_{i,j,k}=\left(\vec{I}_k \times \vec{I}_i\right)\left(\vec{I}_i \cdot \vec{I}_j\right)$ and
\begin{equation}\label{mamalo}
\bar{\mu}(i,j,k)= - \frac{1}{2}\left[ \epsilon^{\mu \nu \rho} D_{i \text{ } \mu x} D_{j \text{ } \nu x} D_{k \text{ } \rho x} + \left( T_i^{[\mu x \text{ } \nu y]} D_{j \text{ } \mu x} D_{k \text{ } \nu y} +  T_j^{[\mu x \text{ } \nu y]} D_{i \text{ } \nu y} D_{k \text{ } \mu x} + T_k^{[\mu x \text{ } \nu y]} D_{i \text{ } \mu x} D_{j \text{ } \nu y} \right) \right],
\end{equation}
with $D_{i \text{ } a} = D_{i \textbf{ } \mu x}$ as in equation (\ref{od}). Note that (\ref{mamalo}) coincides with (\ref{mli}) so Eq. (\ref{milnoAbe}) demands that the third Milnor's coefficient vanishes as required, this is, the next order invariant is meaningful whenever the preceding one vanishes. The consistency conditions (\ref{ccsoa}) and (\ref{milnoAbe}) are also related to a gauge symmetry of the theory. A direct calculation shows that the action (\ref{aim}) is invariant under transformations.
\begin{equation}
    \lambda_{i \text{ } \mu}(x) \enskip \longrightarrow \enskip \lambda_{i \text{ } \mu}(x) + \partial_\mu \omega_i(x), 
    \label{gt2}
\end{equation}
as long as the consistency relations (\ref{ccsoa}) and (\ref{milnoAbe}) are satisfied. Therefore, we observe that both sets of fields $\lambda_i$ and $\Lambda^i$ must be Abelian gauge fields for the theory to be consistent. Note that we have obtained a result analogous to the one in the work \cite{leal-intermedia} concerning the relationship between the consistency and gauge invariance conditions satisfied by the fields of the intermediate action for the third Milnor coefficient $\bar{\mu}(1,2,3)$.

\section{Geometrical interpretation}\label{gi}

\begin{figure}[H]
    \centering    \includegraphics[width=0.5\textwidth]{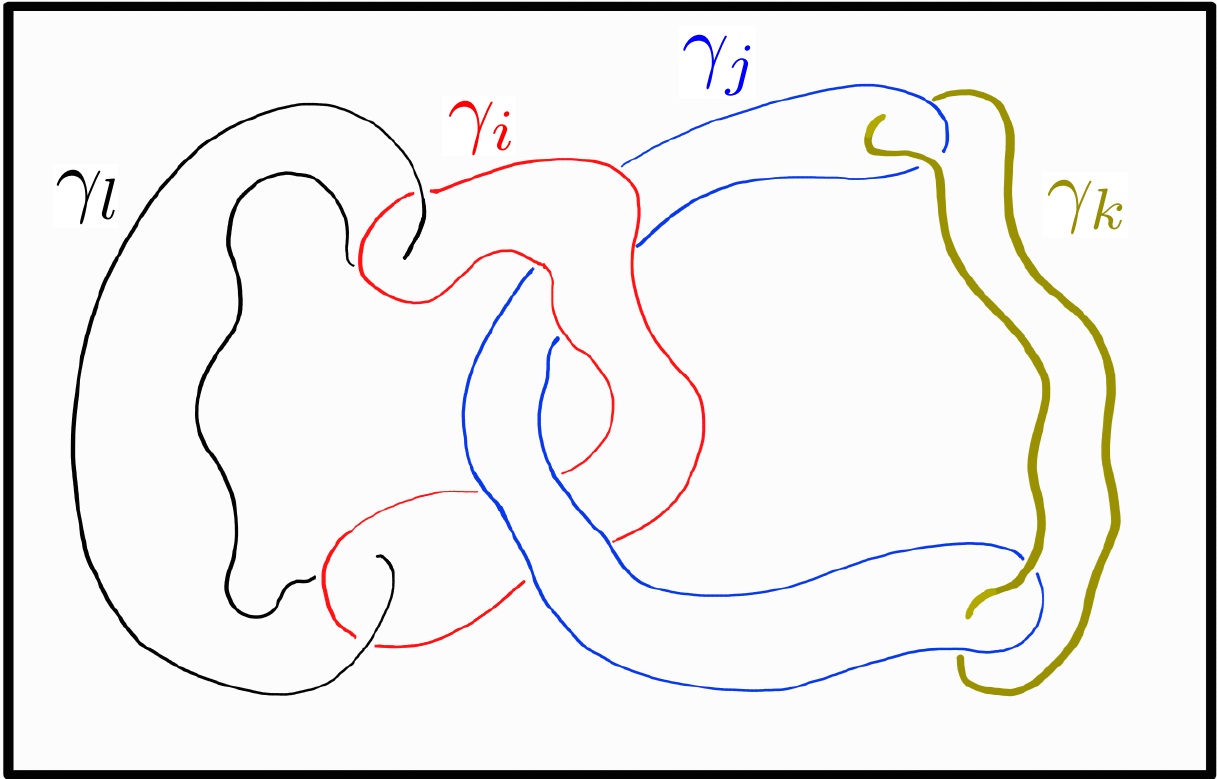}
    \caption{Four component link.}
    \label{l4c5}
\end{figure}

So far, we have found an invariant that we associate with a four-component link by following two paths: through the perturbative development of the CSW action and through an intermediate Abelian theory. In this section, our goal is to demonstrate that indeed, our results correspond to an analytical expression that detects the knotting for the link shown in Fig. \ref{l4c5}
which has vanishing GLN and Third Milnor's coefficient.
Our purpose is to be as clear as possible, so we will begin by rewriting the expressions in a language that we believe is familiar to most of the readers. To this end, we will use the following identity
\begin{equation}
D_{j \textbf{ } \mu x} = \frac{1}{4 \pi} \oint_{\gamma_j} d z^\rho \varepsilon_{\mu \nu \rho} \frac{(x-z)^\nu}{|x-z|^3} \enskip = \enskip \int_{S[\gamma_j, \vec{y}]} d \Sigma_\mu \delta^3{(\vec{x}-\vec{y})} + \partial_\mu f(\vec{x}),
    \label{dsi}
\end{equation}
where $S[\gamma_{j},\vec{y}]$ is the open surface enclosed by the path $\gamma_{j}$, which is parameterized by $\Vec{y}$ and $d\Sigma_{\mu}$ is the normal vector, and $f(\vec{x})$ is a scalar function given by
\begin{equation}
    f(\vec{x})=\frac{1}{4 \pi} \int_{\delta\left[\gamma_j, \vec{y}\right]} d \Sigma_\mu \partial^\mu\left[\frac{1}{|\vec{x}-\vec{y}|}\right].
\end{equation}
Now, the arguments employed throughout this section are constructed using the fields of the intermediate theory $\vec{\lambda}$ and $\vec{\Lambda}$. However, they can be seamlessly applied to the analogous $p$-th order fields of the non-Abelian theory $\Vec{a}^{(p)}$ to analyze higher-order knot invariants.  Let us use the expression (\ref{od}) along with the equation (\ref{eom1}) to rewrite the field $\Vec{\lambda}$ as
\begin{equation}
    \vec{\lambda}_\mu(x) = \frac{1}{2} \sum_{i} D_{i \text{ } \mu x} \vec{I}_i.
    \label{camp0}
\end{equation}
Then, by replacing (\ref{dsi}) into (\ref{camp0}), we have
\begin{equation}
    \vec{\lambda}_\mu(x) = \frac{1}{2} \sum_{i} \left( \int_{S[\gamma_i, \vec{y}]} d \Sigma_\mu \delta^3{(\vec{x}-\vec{y})} \right) \vec{I}_i(0) + \partial_\mu \left( \frac{1}{2} \sum_i f(\vec{x}) \vec{I}_i \right).
    \label{campis}
\end{equation}
Note that, since the theory is gauge invariant, we can set $ \vec{\omega} = - \frac{1}{2} \sum_i f(\vec{x}) \vec{I}_i $ in equation (\ref{gt2}) in such a way that the last term in equation (\ref{campis}) does not appear any more during the computations. For this reason, it is possible to make the following change without any restriction
\begin{equation}
    D_{i \text{ } \mu x} \enskip \longleftrightarrow \enskip \int_{S[\gamma_i, \vec{y}]} d \Sigma_\mu \delta^3{(\vec{x}-\vec{y})}
    \label{Dschg}
\end{equation}

Alternatively, we can find that the following identity 
\begin{eqnarray}\label{eso}
\frac{1}{4 \pi} \frac{(x-z)^{\nu}}{|\vec{x}-\vec{z}|^{3}}=\int_{\gamma^{\vec{z}}}dy^{\nu}\delta^{3}(\vec{y}-\vec{x}) +\varepsilon^{\nu\lambda\rho}\partial_{\lambda}^{\vec{x}} \xi_{\rho}(\vec{x}, \vec{z}),
\end{eqnarray}
holds, if for example, $\gamma^{\vec{z}}$ is as a straight open path coming from infinity to $\vec{z}$, $\xi_{\rho} (\vec{x}, \vec{z})$ is arbitrary, and $\partial_{\lambda}^{\vec{x}}$ is the partial derivative with respect to the $\lambda$-th component of $\vec{x}$. Even more, the presence of $\xi_{\rho}$ makes that $\gamma^{\vec{z}}$ is not unique: it belongs to a family of infinite parallel straight paths ending at some $\vec{z}$ and the curl is a kind of ``gauge'' transformation. Now, contracting equation (\ref{eso}) with $\varepsilon_{\alpha \beta \nu}$ and comparing it with (\ref{g}), we obtain
\begin{equation}
    -g_{\alpha x \text{ } \beta z} = H_{\alpha \beta} (\vec{x}, \gamma^{\vec{z}}) + \partial_\alpha^{\vec{x}} \xi_\beta (\vec{x}, \vec{z}) - \partial_\beta^{\vec{x}} \xi_\alpha (\vec{x}, \vec{z}),
    \label{gheq}
\end{equation}
where
\begin{equation}
    H_{\alpha \beta} (\vec{x}, \gamma^{\vec{z}}) \equiv \int_{\gamma^{\vec{z}}}dy^{\nu} \varepsilon_{\alpha \beta \nu} \delta^{3}(\vec{y}-\vec{x}).
\end{equation}

However, note that terms involving the arbitrary function $\xi_{\alpha}$ in (\ref{gheq}) do not contribute after replacing it in (\ref{aosna2}), as can be easily demonstrated. In this regard, we can consider the following change
\begin{equation}
    -g_{\mu x \text{ } \nu y} \enskip \longleftrightarrow \enskip H_{\mu \nu}\left(\vec{x}, \gamma^{\vec{y}}\right),
    \label{ghchg}
\end{equation}
without lost of generality. In summary, by replacing  (\ref{Dschg}) and (\ref{ghchg}) in  (\ref{aosna2}) we obtain four terms which depends only on closed curves, surfaces and open straight paths that we will analyze in what follows. However, as the contribution of each term will depend on the particular configuration we choose from Figure \ref{l4c5} we will discuss them separately. \\

The first term reads
\begin{equation}
    \begin{split}
        S_A = & -{\int_{\gamma^{\vec{x}_2}} d z^\rho} {\int_{S[\gamma_i, \vec{x_1}]} d \Sigma_\nu} {\int_{S[\gamma_j, \vec{x_2}]} d \Sigma_\beta} {\int_{S[\gamma_k, \vec{x_3}]} d \Sigma_\gamma} {\int_{S[\gamma_l, \vec{x_4}]} d \Sigma_\rho} \varepsilon^{{\nu} {\beta} {\gamma}} \delta^3{({\vec{z}}-{\vec{x}_1})} \delta^3{({\vec{x}_2}-{\vec{x}_3})}  \delta^3{({\vec{z}}-\vec{x_4})}\\ 
        &+ {\int_{\gamma^{\vec{x}_2}} d z^\nu} {\int_{S[\gamma_i, \vec{x_1}]} d \Sigma_\nu} {\int_{S[\gamma_j, \vec{x_2}]} d \Sigma_\beta} {\int_{S[\gamma_k, \vec{x_3}]} d \Sigma_\gamma} {\int_{S[\gamma_l, \vec{x_4}]} d \Sigma_\rho} \varepsilon^{\rho {\beta} {\gamma}} \delta^3{({\vec{z}}-{\vec{x}_1})} \delta^3{({\vec{x}_2}-{\vec{x}_3})}  \delta^3{({\vec{z}}-\vec{x_4})},
    \end{split}
    \label{sa}
\end{equation}
where $\gamma^{\vec{x}_{2}}$ stands for a family of straight open paths from infinity to each point on $\gamma_{j}$.
\begin{figure}[H]
    \centering
    \includegraphics[width=0.5\textwidth]{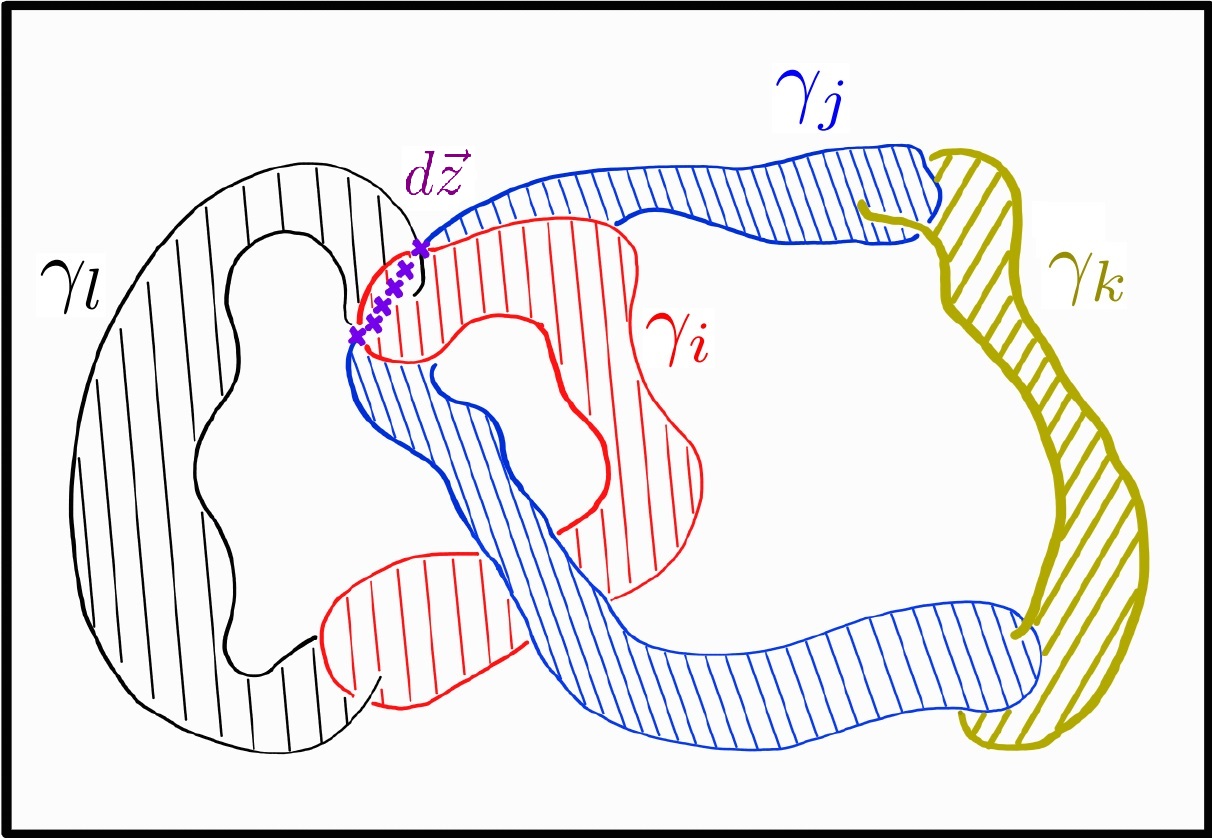}
    \caption{Contribution of the first term in $S^{(2)}$.}
    \label{l4c1}
\end{figure}
Equation (\ref{sa}) takes into account that the surfaces enclosed by  $\gamma_{j}$ and $\gamma_{k}$ (blue and yellow curves/surfaces in Fig. \ref{l4c1}) have points in common, meaning that they intersect each other. Additionally, the contraction of the Levi-Civita symbol with the normal vector of the surfaces enclosed by $\gamma_{i}$, $\gamma_{j}$ and $\gamma_{k}$ (red, blue, and yellow surfaces in Fig. \ref{l4c1}) ensures that the term has non-trivial contribution as those vectors form a non-degenerate volume. Finally, the family of open paths $\gamma^{\vec{x}_{2}}$ must traverse the red and black surfaces before reaching $\gamma_{j}$ (represented by the violets \textcolor{violet}{$\times$}  in the figure) which ensures that $\gamma_{i}$ intersect $S[\gamma_{l},\vec{x}_{1}]$ over the surface $S[\gamma_{j},\vec{x}_{1}]$. More precisely, if
the red and black surfaces in Fig. \ref{l4c1} do not cross over the blue surface, then $\gamma^{\vec{x}_{2}}$ would never cut $S[\gamma_{l},\vec{x}_{1}]$ and $S[\gamma_{i},\vec{x}_{1}]$, and the curves would not be linked. 

The second term can be written as
\begin{equation}
    \begin{split}
        S_B =& - {\int_{\gamma^{\vec{z}_1}} d z^\rho} {\oint_{\gamma_j} d z_1^\nu \int_0^{z_1} d z_2^{\mu_1}} {\int_{S[\gamma_i, \vec{x}_2]} d \Sigma_\nu}  {\int_{S[\gamma_k, \vec{x_3}]} d \Sigma_{\mu_1}}  \int_{S[\gamma_l, \vec{x_4}]} d \Sigma_\rho  \delta^3{({\vec{z}}-{\vec{x}_2})} \delta^3{({\vec{z}_2}-{\vec{x}_3})} \delta^3{({\vec{z}}-\vec{x}_4)}\\
        &+  {\int_{\gamma^{\vec{z}_1}} d z^\nu} {\oint_{\gamma_j} d z_1^\rho \int_0^{z_1} d z_2^{\mu_1}} {\int_{S[\gamma_i, \vec{x}_2]} d \Sigma_\nu}  {\int_{S[\gamma_k, \vec{x_3}]} d \Sigma_{\mu_1}}  \int_{S[\gamma_l, \vec{x_4}]} d \Sigma_\rho  \delta^3{({\vec{z}}-{\vec{x}_2})} \delta^3{({\vec{z}_2}-{\vec{x}_3})} \delta^3{({\vec{z}}-\vec{x}_4)}
    \end{split}
    \label{l4c2}
\end{equation}

\begin{figure}[H]
    \centering
    \includegraphics[width=0.5\textwidth]{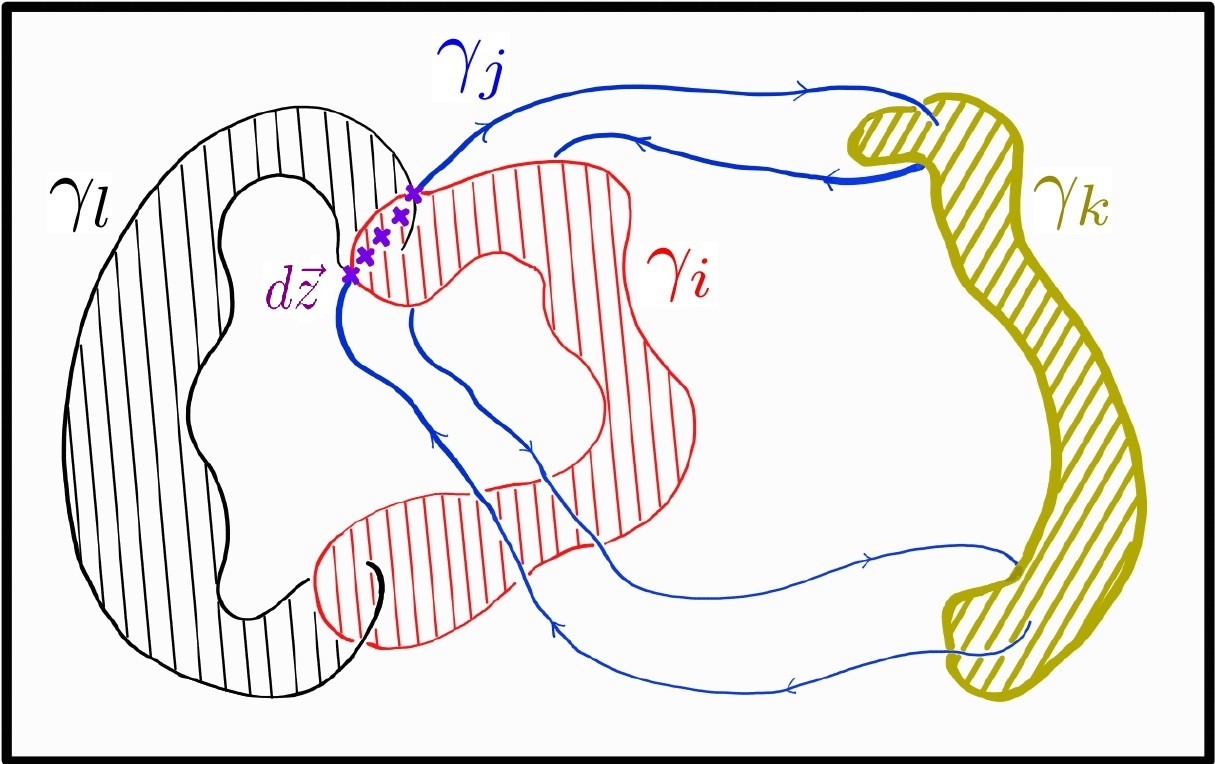}
    \caption{Contribution of the second term in $S^{(2)}$.}
    \label{l4c22}
\end{figure}

and says that $\gamma_{j}$ must intersect the the surface enclosed by $\gamma_{k}$ in one direction and then intersect it in the opposite direction before completely closing the path integral. Once again, terms with path integrals of open curves appear, representing crossings of the red and black surfaces over the blue one similar to the previous case (see Fig. \ref{l4c22}).\\

The third term can be written as
\begin{equation}
    S_C ={ \int_{\gamma^{\vec{z}_3}} d z^\lambda} {\oint_{\gamma_i} d z_1^{\mu} \int_0^{z_1} d z_2^{\mu_1}} {\oint_{\gamma_j} d z_3^{\alpha} \int_0^{z_3} d z_4^{\mu_2}} {\int_{S[\gamma_k, \vec{z}_5]} d \Sigma_{\mu_2}}  \int_{S[\gamma_l, \vec{z}_6]} d \Sigma_{\mu 1} \varepsilon_{{\mu} {\alpha} {\lambda}} \delta^3({\vec{z}}-{\vec{z}_1}) \delta^3{({\vec{z}_4} -{\vec{z}_5})} \delta^3{({\vec{z}_2}-\vec{z}_6)},
    \label{l4c3}
\end{equation}

\begin{figure}[H]
    \centering
    \includegraphics[width=0.5\textwidth]{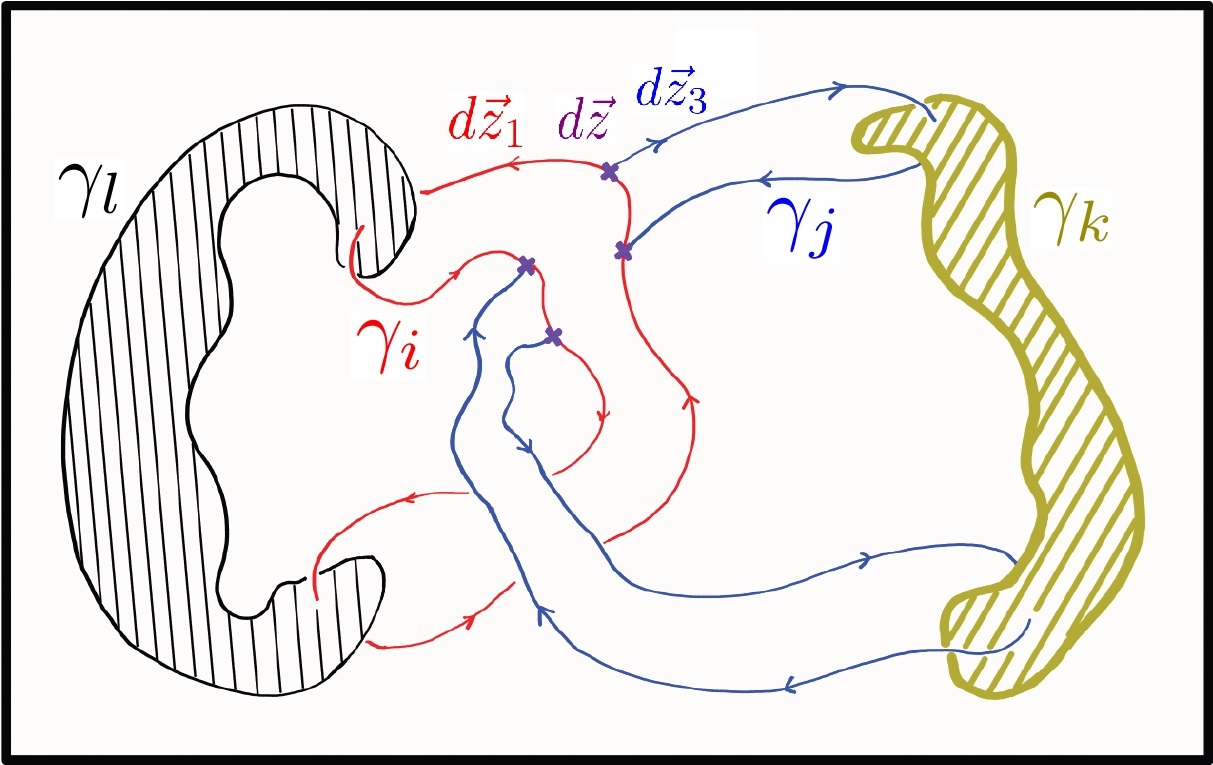}
    \caption{Contribution of the third term in $S^{(2)}$.}
    \label{l4c33}
\end{figure}
and takes into account that $\gamma_{j}$ intersects the surface enclosed by $\gamma_{k}$ in one direction and then intersects it in the opposite direction to close the path integral. Similarly, with $\gamma_{i}$ which intersects the surface enclosed by $\gamma_{l}$ and then intersects it again in the opposite direction to close the integral (see Fig. \ref{l4c33}). Note that we also have an integral on $\gamma^{\vec{z}_{3}}$ ending on $\gamma_{j}$, which cuts through the $\gamma_{i}$, representing the crossing of the red curve over the blue one (because otherwise, there is no knotting). Finally, the tangent vectors to $\gamma_{i}$, $\gamma_{j}$ and the open curve must form a non-degenerate volume, as shown in the figure. Note that the tangent vectors to the open path $\gamma^{\vec{z}_{3}}$ enter the screen and are represented by \textcolor{violet}{$\times$} in Fig. \ref{l4c33}.\\

Finally, the last term is given by
\begin{equation}
    S_D =  - {\oint_{\gamma_j} d z_1^\mu \int_0^{z_1} d z_2^{\mu_1} \int_0^{z_2} d z_3^{\mu_2}} {\int_{S[\gamma_i, \vec{z}_4]} d \Sigma_{\mu_1}}  {\int_{S[\gamma_k, \vec{z}_5]} d \Sigma_{\mu_2}}  \int_{S[\gamma_l, \vec{z}_6]} d \Sigma_{\mu} \delta^3{({\vec{z}_1}-\vec{z}_6)} \delta^3{({\vec{z}_2}-{\vec{z}_4})} \delta^3{({\vec{z}_3}-{\vec{z}_5})}.
    \label{l4c4}
\end{equation}

\begin{figure}[H]
    \centering
    \includegraphics[width=0.4\textwidth]{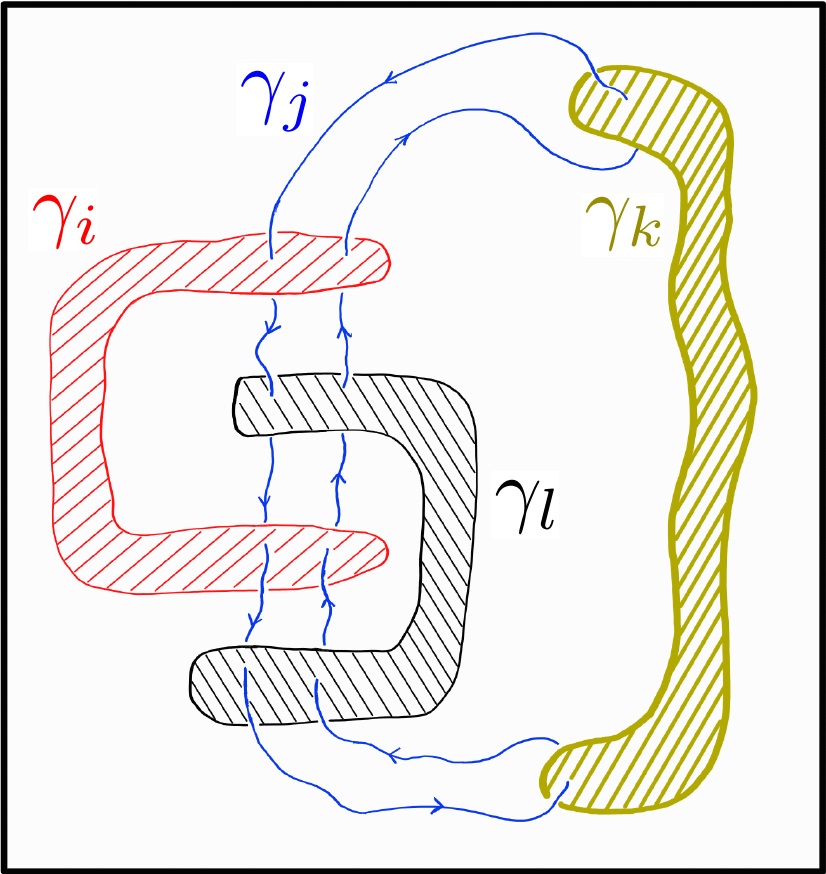}
    \caption{Contribution of the fourth term in $S^{(2)}$.}
    \label{l4c4u}
\end{figure}
which takes into account that $\gamma_{j}$ must intersect the surface enclosed by $\gamma_{k}$, then intersect the surface enclosed by $\gamma_{i}$, and finally the surface enclosed by $\gamma_{l}$. Now, since the outermost integral is a closed path integral, it must return to the initial point. To achieve this, it must intersect $\gamma_{k}$ again in the opposite direction. However, as the two subsequent inner integrals are open integrals that precede the closed path integral, $\gamma_{j}$ must first intersect in the opposite direction to the surfaces enclosed by $\gamma_{l}$ and $\gamma_{i}$ before reaching the initial point, thus closing the path integral as shown in the Fig. \ref{l4c4u}. Note that this particular term can be seen as a thread crossing over ``three fabrics'' of different colors; in other words, the integral in (\ref{l4c4}) tells us how to weave the blue curve over the other surfaces. \\

To conclude this section, we would like to emphasize that the analysis we have shown here, has been performed for specific configurations from Figure \ref{l4c5}. Consequently, not all terms contribute simultaneously. For example, note that the configurations in Figs. \ref{l4c1} and \ref{l4c22} are equivalent, and therefore, the terms $S_{A}$ and $S_{B}$ contribute to the linking. However, based on these figures, the terms $S_{C}$ and $S_{D}$ do not contribute. Similarly, considering the configuration in Fig. \ref{l4c33}, the only contributing term to the linking is $S_{C}$, and for Fig. \ref{l4c4u}, all terms vanish except $S_{D}$. In any case, it is observed that regardless of the configuration, the action (\ref{aosna2}) detects the linking of the four-component link, as expected.

\section{conclusions}\label{conclusions}

In this work, we develop into the Chern-Simons-Wong theory and its connection with knot invariants based on a previous work where it was demonstrated that the Chern-Simons-Wong action, being a topological quantity, carries information about the entanglement of world lines associated with particles. We introduced a method that yields invariant functions from classical field theories. In the
perturbative regime we actually obtain a countable hierarchy of invariants. Our procedure provides explicit formulae for the higher
order invariants, and our computations show that the consistency equations are satisfied. We expect the higher order link invariants arising in the computation
of the perturbative Chern-Simons-Wong on-shell action, to be closely related to Milnor’s link invariants \cite{milnor}.
\\

We believe our methods can be usefully applied to other classical field theories. In particular, it may be rewarding to look at Yang-Mills-Wong action in higher dimensions, it should yield conformal invariants associated with closed curves in spacetime. It may also be interesting to apply our methods to the generalized Chern-Simons action \cite{schwar}, it should yield invariants related with Chas-Sullivan product in string topology \cite{cattaneo,chas}. \\

The main contribution of this work is the explicit form of the second-order action, revealing a direct correlation with the invariant of a four-component link. Furthermore, an Abelian intermediate model was constructed, reproducing the same invariant, highlighting the robustness and versatility of the findings.
\\

In our study of perturbative solutions we saw that the invertibility of the quadratic part of the action plays a fundamental role. Often the quadratic part is not invertible and new techniques are required in order to get invariants. One possibility is to introduce, as in the quantum case, fermionic variables and replace the action with a new one with invertible quadratic part. Thus, it is plausible that in the classical perturbative regime, the BRST and BV procedures may still play a role. Another possibility arises when the inverse of the quadratic part of the action is no quite well-defined, but rather a singular operator. In this case techniques from renormalization \cite{connes} may become useful in order to replace invariants given by ill-defined divergent integrals, by their renormalized values.\\

The geometric interpretation of the on-shell action provides strong evidence connecting it to the linking of the four-component link. It is worth mentioning that the four-component link appears to be associated with the entanglement of a two-component link when considering degenerate currents. The invariant involving two curves, distinct from the Gauss linking number (consistently zero in this theory), is proposed to be connected to the Whitehead link. This could be an immediate extension for this work, first characterizing the second order action so that it contemplates only two components of closed curves and then applying the interpretive scheme to verify that certainly it thakes in account the linking properties of this other invariant, that seems to be at the same level as that of four components \cite{mona}. It has become clear that many constructions in field theory, as well as in other branches of physics and mathematics \cite{categora1,categora2,categora3} admit categorical analogues. It would be interesting to investigate the categorical foundations of the method introduced in this paper.\\

To finish the discussion we will mention that it seems natural to accept that successive corrections to the action (evaluated over the equations of motion of non-Abelian particles) can be identified, and provide analytical expressions, for higher order linking coefficients. The main problem lies in the fact that as we increase the order in the perturbative expansion, the complexity and cumbersomeness of the calculation (size of the expressions) grows considerably, making it no longer manageable. For this reason, it is of vital importance to propose a diagram (in the style of Feynman diagrams) that allows directly obtaining the successive corrections and thus the possible detection of their associated invariants.  However, exploration of these aspects extends beyond the scope of this study, suggesting promising avenues for future research.

\end{document}